\documentclass[prb,twocolumn,amsmath,showpacs]{revtex4}

\usepackage{graphicx}

\begin{document}
\input epsf

\title {On the equilibrium magnetization of high-$T_{c}$ superconductors 
below the irreversibility line}

\author {I. L. Landau$^{1,2}$ H. R. Ott$^{1}$}
\affiliation{$^{1}$Laboratorium f\"ur Festk\"orperphysik, ETH H\"onggerberg, 
CH-8093 Z\"urich, Switzerland}
\affiliation{$^{2}$Kapitza Institute for Physical Problems, 117334 Moscow, 
Russia}

\date{\today}

\begin{abstract}
By scaling isothermal magnetization data measured at different temperatures 
in the mixed state of high-$T_{c}$ superconductors, we show that in some 
cases the sample magnetization, measured in increasing magnetic field below 
the irreversibility line, is identical with the equilibrium magnetization 
even in magnetic fields well within the irreversible regime. This surprising 
behavior can hardly be explained in terms of traditional models of vortex 
pinning in the bulk of the sample. 
\end{abstract}
\pacs{74.25.Op, 74.25.Qt, 74.72.-h}

\maketitle

One of the specific features of the field-induced magnetization in 
high-$T_{c}$ superconductors (HTSC's) is that there is an extended range 
of external magnetic fields $H$ below the upper critical field $H_{c2}$, 
where the sample magnetization $M$ is reversible, \cite{muller} i.e., the 
values of $M$ measured in either increasing or decreasing magnetic fields 
coincide. The lower boundary of this range is the so-called irreversibility 
line (IRL) in the $H$-$T$ phase diagram and the values of $M$ measured 
above the IRL represent the equilibrium magnetization $M_{eq}$. There is 
no reliable way to evaluate $M_{eq}$ from the experimental data below the 
IRL without some additional knowledge about the pinning mechanisms in the 
particular sample under investigation. Although different varieties of 
the critical state model are often used for the analysis of experimental 
results, their applicability is very rarely justified and therefore, the 
results of those analyses are not reliable. For instance, the simplest 
and most widely used critical-state model of Bean is based on the 
assumption that the critical current density $j_{c}$ is independent of 
the magnetic induction. \cite{bean1,bean2} Experiments show, however, that 
$j_{c}$ in HTSC's strongly depends on the applied magnetic field, i.e., 
the Bean model is not really valid for describing the critical state 
of these materials. It has also been demonstrated that the ÒequilibriumÓ 
magnetization curves derived from magnetization data obtained below the 
IRL by employing the Bean model do not really represent $M_{eq}$. 
\cite{phys} In this work, we demonstrate how a scaling procedure, recently 
developed in Ref. \onlinecite{prb}, may successfully be used for the 
analysis of experimental magnetization $M(H)$ curves below the IRL and 
how, as a consequence, important information concerning the effective 
pinning of vortices may be obtained. 

The scaling procedure is based on the single assumption that the 
Ginzburg-Landau (GL) parameter $\kappa$ is temperature independent. In 
this case, the magnetic susceptibility of a superconductor in the mixed 
state $\chi (H,T)$ is a universal function of $H/H_{c2}(T)$ and the relation 
between the magnetizations at two different temperatures $T$ and $T_{0}$ 
may be written as
\begin{equation}
M(H/h_{c2},T_0)=M(H,T)/h_{c2}
\end{equation}
with $h_{c2} = H_{c2}(T)/H_{c2}(T_{0})$ being the normalized upper 
critical field. Considering real HTSC's, we also have to take into 
account the temperature dependent paramagnetic susceptibility $\chi_{n}$ 
of the normal vortex cores which, according to Ref. \onlinecite{prb}, 
leads to the relation
\begin{equation}
M_{eff}(H/h_{c2},T_0)=M(H,T)/h_{c2}-c_0(T)H
\end{equation}
with $c_{0}(T) = \chi_{n} (T_{0}) - \chi_{n} (T)$. Eq. (2) implies that 
the field dependence of the sample magnetization $M(H)$ at a chosen 
temperature $T_{0}$ may be obtained from $M(H)$ curves measured at 
different temperatures. The collapse of these individual $M(H)$ curves 
onto a single master curve  may be achieved by a suitable choice of 
$h_{c2}(T)$ and $c_{0}(T)$, the adjustable parameters of the scaling 
procedure. The scaling procedure is only applicable to magnetization data 
collected above the IRL. In this case, $M_{eff}(H) = M_{eq}(H,T_{0})$. 
At the same time, once $h_{c2}(T)$ and $c_{0}(T)$ have been established 
in the chosen range of temperatures, the transformation given by Eq. (2) 
may also be applied to magnetization data measured below the IRL. Because 
of the onset of irreversibility, $M_{eff}(H,T_{0})$ generally no longer 
represents $M_{eq}(H,T_{0})$. However, as will be shown below, a surprising 
asymmetry of the $M_{eff}$ curves, calculated from $M(H)$ data taken in 
increasing and decreasing fields, with respect to the equilibrium 
magnetization curve offers to achieve important conclusions concerning 
the effective pinning mechanism.

The condition that $\chi (H,T)$ depends only on the ratio $H/H_{c2}(T)$, 
which is the essential background of the scaling procedure, remains valid 
for any configuration of the mixed state. The vortices may form a vortex 
lattice, a vortex liquid, or, as has recently been proposed, a system of 
superconducting filaments embedded in the matrix of the normal metal. 
\cite{model} This circumstance provides the possibility to use the 
scaling procedure even if there is a step in the $M(H)$ curves, 
marking the so-called first order phase transition in the mixed 
state of HTSC's, which usually is attributed to the melting of the vortex 
lattice. \cite{tr1, tr2, tr3, tr4, tr5} Although the vortex lattice 
melting represents a rather plausible hypothesis, to the best of our 
knowledge, there are no direct experimental evidences for this claim. For 
our discussion, however, the real nature of the phase transition does not 
need to be known. It is only  important that in the $H$-$T$ phase diagram 
there is a boundary $H_{PT}(T)$ between two possible configurations of 
the mixed state. In this case, at a fixed temperature and with increasing 
magnetic field, a phase transition leads from one configuration (low-field 
phase) to the other (high-field phase). By $M_{eq}^{(l)}(H)$ and 
$M_{eq}^{(h)}(H)$ we denote the equilibrium field-induced variations of 
$M$ in the low-field and high-field phase, respectively. An example is 
shown in the inset of Fig. 1(a). \cite{ftnt} Of course, $M_{eq}^{(l)}(H)$ 
and $M_{eq}^{(h)}(H)$ do not coincide, but they both should scale with 
the same values of $h_{c2}(T)$ and $c_{0}(T)$. In this work we concentrate 
on the features of the magnetization curves distinctly above and below 
the phase transition. A detailed analysis of the magnetization very close 
to the phase transition will be published elsewhere. 
\begin{figure}[t]
 \begin{center}
  \epsfxsize=1\columnwidth \epsfbox {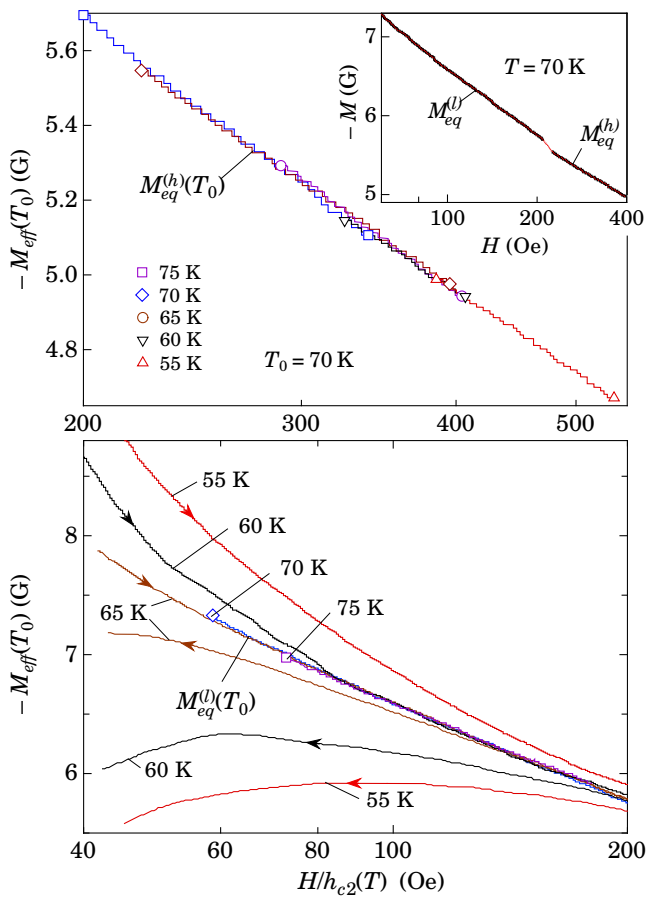}
  \caption{$M_{eff}(H,T_{0})$ for sample Bi-1 (original $M(H)$ data 
           taken from Ref. \onlinecite{kimura1}), (a) above and (b) below 
           the first order transition. The open symbols mark the end points 
           of the covered field ranges at the indicated temperatures. The 
           inset illustrates the definitions of 
           $M_{eq}^{(h)}$ and $M_{eq}^{(l)}$, taking the $M(H)$ curve at 
           $T=70$ K as an example. The $M_{eff}(H,T_{0})$ curves were 
           calculated using Eq. (2) with $T_{0} = 70$ K (see text).}
 \end{center}
\end{figure}

Below we consider results of the magnetization measurements for three 
Bi$_{2}$Sr$_{2}$CaCu$_{2}$O$_{8+x}$ (Bi-1, Bi-2, and Bi-3) single 
crystals that were reported in Refs. \onlinecite{kimura1,kimura2,kado}, 
respectively. In all three cases only the magnetization data from above 
the IRL were used to establish the parameters $h_{c2}(T)$ and $c_{0}(T)$. 

Figure 1 shows $M_{eff}(H,T_{0})$ data for the sample Bi-1.  At $T \ge 
60$ K the IRL line for this sample is substantially below $H_{PT}(T)$. 
\cite{kimura1} The scaled magnetization curves above the phase 
transition are depicted in Fig. 1(a). Because these data were collected 
above the IRL, the resulting curve in Fig. 1(a) represents the 
equilibrium $M_{eq}^{(l)}(H)$ curve for $T_{0} = 70$ K. The 
magnetization data collected below the phase transition are shown in 
Fig. 1(b). The magnetization of this sample measured at temperatures of 
70 K and 75 K is reversible in the entire covered range of fields and 
therefore, the corresponding curves in Fig. 1(b) represent the 
$M_{eq}^{(l)}(H,T_{0})$ curve. The merging of the individual 
$M(H)$ curves to $M_{eq}^{(h)}(H,T_{0})$ and $M_{eq}^{(l)}(H,T_{0})$, as 
displayed in Figs. 1(a) and 1(b), was achieved with the same values of 
$h_{c2}(T)$ and $c_{0}(T)$ on the both sides of the transition, thus 
confirming our claim above. Although the magnetization at $T \le 65$ 
K is irreversible in low magnetic fields, the $M_{eff}(H)$ curves 
calculated from the magnetization data measured in increasing magnetic 
field at 60 and 65 K merge into the equilibrium magnetization curve in 
magnetic fields considerably below the corresponding values of the 
irreversibility field $H_{irr}$. This is obviously not the case for 
$M_{eff}(H,T_{0})$ calculated  from $M(H)$ data taken in decreasing field, 
revealing an asymmetry of the magnetization process.
\begin{figure}[h]
 \begin{center}
  \epsfxsize=1\columnwidth \epsfbox {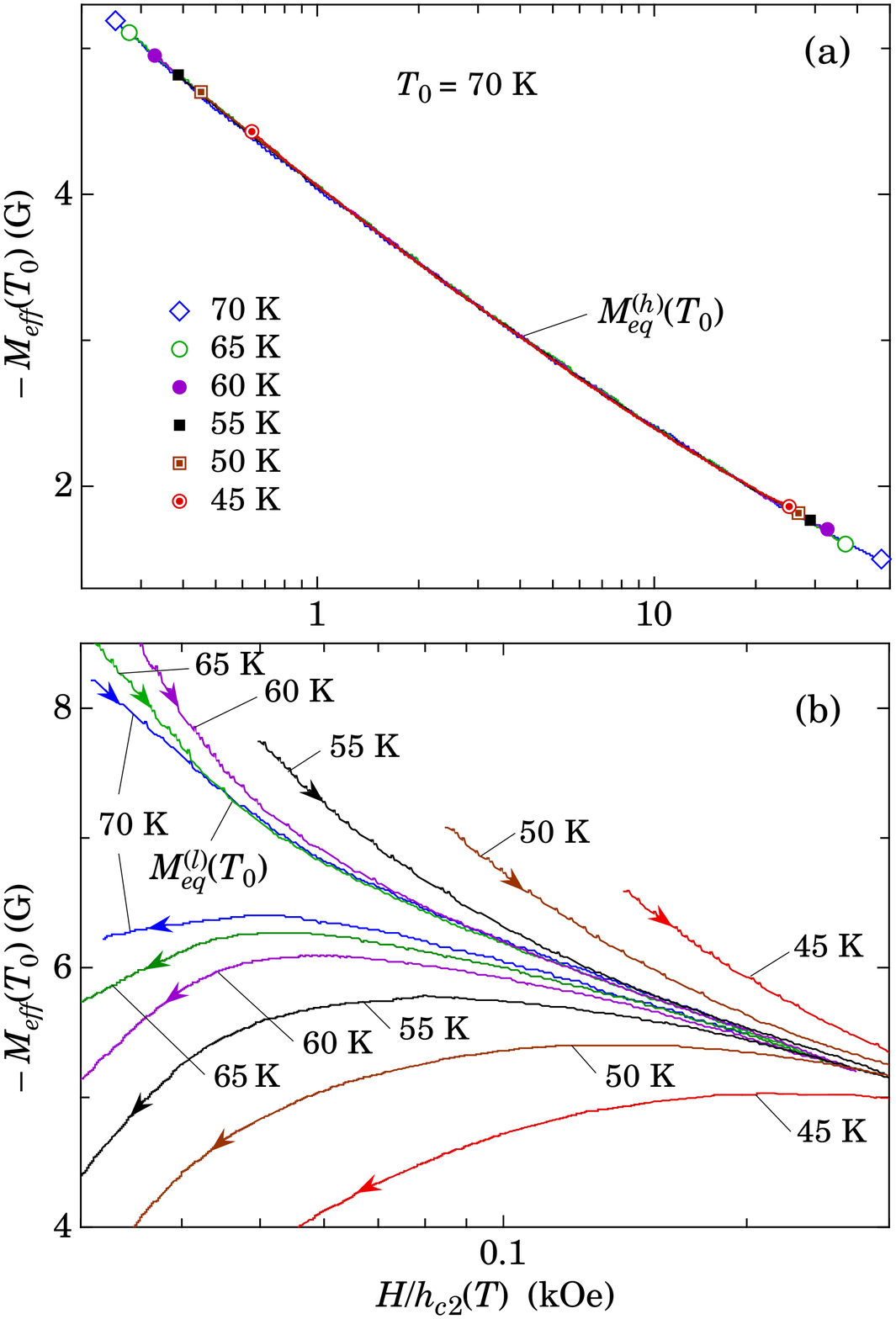}
  \caption{$M_{eff}(H,$70K) for sample Bi-2 (original data taken from 
           Ref. \onlinecite{kimura2}), (a) above and (b) below the phase 
           transition. The  $M_{eff}(H,T_{0})$ curves were calculated 
           using Eq. (2).} 
 \end{center}
\end{figure}

Analogous results for the sample Bi-2 are shown in Fig. 2. In contrast 
to the previous case, $H_{irr}(T)$ for sample Bi-2 is practically 
identical with $H_{PT}(T)$, marking the phase transition, at all 
temperatures. \cite{kimura2} Because the relative magnetic field range 
covered in Ref. \onlinecite{kimura2} is extremely wide, accurate and 
reliable values of $h_{c2}(T)$ and $c_{0}(T)$ were obtained. As is 
demonstrated in Fig. 2(a), the scaling procedure results in a perfect 
overlap of the $M(H)$ curves above the IRL and deviations between the 
data measured at different temperatures are of the order of the width 
of the line. Fig. 2(b) emphasizes the features of $M_{eff}(H,T_{0})$ 
below the transition. Similar to the previous case, the $M_{eff}(H,T_{0})$ 
curves calculated from the measurements in increasing field at $T \ge 55$ 
K coincide already in a magnetic field range which extends to substantially 
below $H_{irr}$. This is only possible if each of the coinciding parts 
of the curves is calculated from the equilibrium magnetizations at the 
respective temperatures. Again, due to irreversibility the $M_{eff}(H)$ 
curves deviate from the equilibrium magnetization curve at lower 
temperatures, but these deviations are again noticeably smaller for the 
measurements made in increasing field than for those made with decreasing 
field. 

A third set of data is shown in Fig. 3. For this plot we have chosen 
only the $M(H)$ data measured at several temperatures rather close to 
the critical temperature $T_{c}$. Although the first order phase 
transition clearly manifests itself on the magnetization curves at 
lower temperatures, \cite{kado} it is practically invisible in this 
high temperature range. As may be seen in Fig. 3, the 
$M_{eff}(H)$ curves, calculated from the measurements in increasing 
field, all merge in the entire covered ranges of fields, thus 
clearly indicating that this curve represents the equilibrium 
magnetization curve for $T=T_{0}$.
\begin{figure}[h]
 \begin{center}
  \epsfxsize=1\columnwidth \epsfbox {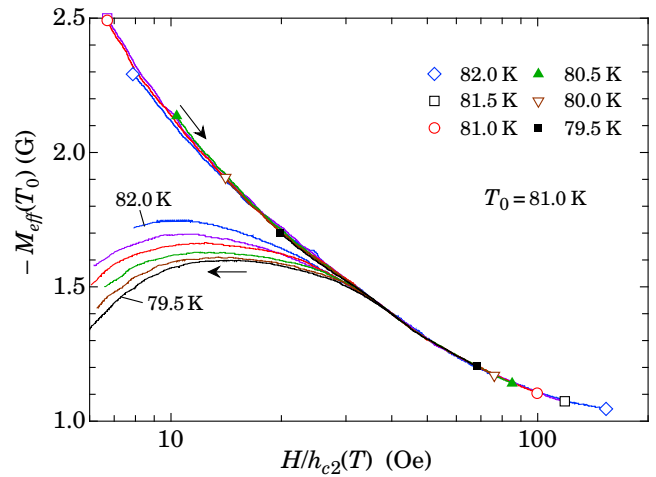}
  \caption{$M_{eff}(H,T_{0})$ curves calculated using Eq. (2) with 
           $T_{0} = 81$ K (see text) for sample Bi-3. Original $M(H)$ 
           data are taken from Ref. \onlinecite{kado}. The symbols 
           mark the end points of the covered field ranges at the 
           indicated temperatures.}
 \end{center}
\end{figure}

The data shown in Figs. 1-3 demonstrate that the effect of pinning is 
strongly dependent on the direction of the flux motion. The pinning 
effects are obviously much weaker for the magnetic flux entering 
the sample. We are not aware of any model that explains this kind of 
pinning force asymmetry, if these forces are related to pinning centers 
in the bulk of the sample. A reasonable explanation for this type of 
behavior might be, however, that in these high quality samples, the 
intrinsic pinning is weak and the main obstacle for the magnetic-flux 
motion is a barrier near the sample edges, the so-called geometrical 
barrier. The existence of this type of barriers is actually known 
since the early studies of the intermediate state in type-I 
superconductors employing a magnetic powder technique. \cite{shar1} 
These experiments have shown that the concentration of the normal 
phase in the intermediate state of type-I superconductors is 
considerably smaller near the sample edges. \cite{shar1ad}. It was 
immediately recognized that this happens because of the non-ellipsoidal 
shape of the sample. Indeed, as is well known, if the magnetic 
susceptibility $\chi$ is nonzero, the magnetic induction $B$ is uniform 
only in ellipsoidal samples. \cite{ellips} In superconducting samples 
this non-uniformity of $B$ is magnified by a strong dependence of $\chi$ 
on the magnetic induction. The resulting distribution of shielding 
currents effectively pushes the normal domains in the intermediate state 
of type-I superconductors as well as vortices in the mixed state of 
type-II superconductors towards the center of the sample. It was also 
demonstrated that this edge barrier for the flux motion in type-I 
superconductors may substantially be reduced by proper shielding of the 
sample edges \cite{shar2} or by altering the sample shape. \cite{castro} 
The importance of this edge barrier for correct interpretations of 
experimental data was also recognized for HTSC's. \cite{geo1,geo2,geo3} 

The geometrical barrier reaches its maximum height very close to the 
sample edges and the corresponding potential decreases only gradually 
towards the center of the sample. \cite{geo1} This asymmetry of the 
potential profile implies the corresponding asymmetry of its effect on 
the vortex motion. The geometrical barrier naturally represents a 
stronger obstacle for the vortex motion out of the sample because it 
keeps the vortices at some considerable distance from the sample 
edges and therefore, thermal activation is ineffective for the exit of 
vortices. Because of the proximity of the potential maximum to the 
sample edges, the thermally activated entrance of vortices is much more 
likely than their exit. This simple model explains why the data presented 
in Figs. 1-3 are consistent with the assumption that the pinning in the 
bulk of the sample is negligible and that the irreversibility of the 
magnetization is due to the mentioned geometrical barrier. Because the 
height of the geometrical barrier is strongly dependent on the shape of 
the sample edges, it may vary significantly from sample to sample.

In many experimental studies, including that of Ref. \onlinecite{kimura2}, 
the irreversibility line in the $H$-$T$ phase diagram practically coincides 
with the line marking the first order phase transition. The standard 
interpretation of this onset of irreversibility rests on the nonzero shear 
modulus of the vortex lattice, causing it to be much stronger pinned than 
the vortex liquid. This may well be true for the bulk pinning, but the 
shear modulus of the vortex lattice is irrelevant for the entry or exit 
of the vortices across the geometrical barrier. In other words, if the 
bulk pinning is weak compared to the pinning arising from the sample edges, 
which seems to be the case at least for our three examples and possibly 
many other HTSC's, the onset of the irreversibility at the mentioned phase 
transition does not necessarily follow from postulating the melting of the 
vortex lattice. 

With all this in mind, we suggest an alternative cause for the occurrence 
of the first order phase transition in the mixed state of HTSC's. As was 
argued in Ref. \onlinecite{model}, it seems possible that in high enough 
magnetic fields, the mixed state is formed by a system of superconducting 
filaments embedded in the matrix of the normal metal instead of the 
formation of Abrikosov vortices. Upon reducing the magnetic field, the 
system of superconducting filaments looses its stability and must undergo 
a transition to the traditional mixed state consisting of Abrikosov 
vortices in a superconducting matrix. This transition requires a complete 
change of the topology of the system and, although it is not exactly a 
first order phase transition, its principal features will include the 
occurrence of a latent heat, a discontinuity in the magnetization, and 
hysteresis effects. It should be noted that the sample magnetization for 
a mixed state consisting of superconducting filaments is always reversible, 
independent of whether the filaments are pinned or not. In this case, of 
course, the geometrical barrier has no influence on the reversibility of 
the sample magnetization. The transition to the traditional mixed state 
with Abrikosov vortices changes this situation completely and, if the 
vortices are pinned, $H_{irr}$ naturally coincides with the phase 
transition. \cite{comm2} The same is true with respect to the sample 
resistivity. It is clear that, because there is no direct superconducting 
link between one electrode and the other for the system of superconducting 
filaments, the sample resistivity should drop with the transition to 
Abrikosov vortices. The magnitude of this resistance jump depends on the 
strength of the vortex pinning and, for strong pinning, the sample 
resistance may vanish at the transition point. 

As demonstrated above, in a number of cases $M(H)$ measured in increasing 
magnetic field coincides with the equilibrium magnetization curve even 
in magnetic fields well below the IRL, which is a strong evidence that 
the geometrical barrier arising near the sample edges is the main obstacle 
for the motion of magnetic flux. If this is indeed the case, the onset of 
irreversibility and the resistivity jump at the transition point do not 
necessarily follow from the hypothesis of the vortex lattice melting. In 
this sense we also promote an alternative scenario for explaining the first 
order phase transition in the mixed state of HTSC's.


\begin{thebibliography}{71}
\bibitem{ftnt} In HTSC samples, in which the first-order phase transition 
is observed, the IRL in the $H$-$T$ phase diagram either coincides with 
the $H_{PT}(T)$ curve or lies below it, apparently depending on the sampe 
quality. Only if $H_{irr}(T) < H_{PT}(T)$, as it is the case for the data 
shown in the inset of Fig. 1(a), $M_{eq}^{(l)}(H)$ may be derived directly 
from experimental measurements.
\bibitem{shar1ad} These observations, unfortunately, have never been 
published and only a brief discussion on this matter may be found in Ref. 
\onlinecite{shar2}. Illustrative images of this effect, however, are 
given in Ref. \onlinecite{castro}.
\bibitem{comm2} In samples with weak vortex pinning, the IRL may, of 
course, be shifted such that $H_{irr}(T) < H_{PT}(T)$.
\end{thebibliography}
\end{document}